\documentclass [11pt]{article}
\usepackage[dvips]{graphicx}
\usepackage{subcaption}

\usepackage{epsfig}
\usepackage{graphicx}
\usepackage{amsmath}
\usepackage{amsfonts}
\usepackage{amssymb}
\usepackage{hyperref}
\usepackage{latexsym}

\usepackage{color}
\usepackage{hyperref}

\begin{document}

\begin{center}
\begin{flushright}
	\begin{small}    
\end{small}
 \end{flushright} \vspace{1.5cm}
\Large{\bf Inflationary Scenarios in $f(Q,\phi)$ Gravity with Scalar Field Coupling }
\end{center}

\begin{center}
F. Mavoa $^{(a,b)}$\footnote{e-mail: ferdinand.mavoa@univ-labe.edu.gn, maferdson@yahoo.fr},
M.B. Barry $^{(a)}$\footnote{mamadou-bailo.barry@univ-labe.edu.gn},
R.Ndioukane $^{(a)}$\footnote{ remindioukane@gmail.com},
M. G. Ganiou $^{(c)}$\footnote{ganiou.gbenga@uganc.edu.gn}, 
F.K. Ahloui $^{(d)}$\footnote{ahlouiflorentin@yahoo.fr}

\vskip 4mm
$^{a}$\, {\it  D\'epartement Energie Photovolta\"{i}que, Universit\'e de Lab\'e, R\'epublique de Guin\'ee}\\ 

{\it B.P:(+224) 2010,  Lab\'e, R\'epublique de Guin\'ee}\\

$^b$ \,{\it International Chair of Mathematical Physics and Application (ICMPA) , University d'Abomey-Calavi, B\'enin}\\
 {\it  072 BP 50, Cotonou, B\'enin}\\

$^{c}$\, {\it  D\'epartementde Phsique , Universit\'e Gamal Abdel Nasser, R\'epublique de Guin\'ee}\\
{\it BP 1147 Conakry , R\'epublique de Guin\'ee}\\
 
$^{d}$\, {\it  Institut Superieur des sciences de l'Education,  Conakry}\\
{\it BP 1147 Conakry  , R\'epublique  Guin\'ee}\\

\vskip 2mm
\end{center}

\begin{abstract}


In this work, we investigated several inflationary scenarios within the framework of modified $f(Q,\phi)$ gravity with a nonminimal coupling between the scalar field and the nonmetricity scalar. We focused on the impact of the coupling parameter $\xi$ on the inflationary observables, namely the scalar spectral index $n_s$ and the tensor-to-scalar ratio $r$. In the case of De Sitter inflation, we showed that the model can reproduce observationally viable predictions only within a restricted range of the coupling parameter. Specifically, we found that $n_s$ increases with $\xi$, while $r$ decreases, leading to a narrow allowed region $10^{-3} \lesssim \xi \lesssim 10^{-2}$ compatible with Planck data. Outside this range, the model either predicts excessively large tensor modes or an unphysical blue-tilted spectrum. We also derived theoretical constraints on $\xi$ from the consistency of the model, leading to an upper bound $\xi < \frac{\kappa}{2p}$. For $\kappa = 1$ and $p = 60$, this implies $\xi < 0.00833$, with a preferred region around $\xi \sim \mathcal{O}(10^{-3})$. Furthermore, we analyzed the Cosh-type inflationary model and showed that it provides a robust and consistent description of inflation. In this case, the tensor-to-scalar ratio decreases while the scalar spectral index increases with the number of e-folds $N$. For $N = 60$, the model predicts $n_s \approx 0.965 - 0.967, \qquad r \approx 0.017 - 0.018$, in excellent agreement with current observational constraints. Overall, our results highlight the crucial role of the nonminimal coupling in shaping the inflationary dynamics and ensuring compatibility with cosmological observations.
\hspace{0,2cm} 
        
\end{abstract}


\section{Introduction}

	Over the past few decades, significant progress in observational astronomy has profoundly improved our understanding of the Universe. It is now well established that the Universe is not only expanding but also undergoing an accelerated expansion. This remarkable discovery was first revealed through Type Ia supernova observations by \textit{Riess et al.}~\cite{Riess1998, Riess2004} and \textit{Perlmutter et al.}~\cite{Perlmutter1999}, and later supported by baryon acoustic oscillations \cite{Eisenstein2005} and cosmic microwave background measurements from the \textit{Planck Collaboration}~\cite{Planck2020}. Consequently, explaining this accelerated expansion has become a central issue in modern cosmology.
	
	Within the standard cosmological model, this behavior is attributed to dark energy, which constitutes about $70\%$ of the total energy density of the Universe. In the $\Lambda$CDM framework, dark energy is modeled as a cosmological constant. However, this simple picture faces severe theoretical difficulties, notably the fine-tuning and coincidence problems discussed by \textit{Weinberg}~\cite{Weinberg1989} and \textit{Peebles \& Ratra}~\cite{PeeblesRatra2003}.
	
	To address these issues, several modifications of General Relativity have been proposed. One of the earliest and most studied extensions is $f(R)$ gravity, developed and reviewed by \textit{Sotiriou \& Faraoni}~\cite{SotiriouFaraoni2010}. This framework has been generalized into scalar–tensor theories as described by \textit{Fujii \& Maeda}~\cite{FujiiMaeda2007}, as well as Gauss–Bonnet modifications introduced by \textit{Nojiri \& Odintsov}~\cite{NojiriOdintsov2005}. These theories introduce additional dynamical degrees of freedom through scalar fields or higher-order curvature corrections.
	
	However, curvature-based modifications often lead to higher-order field equations, as shown by \textit{Stelle}~\cite{Stelle1977}, which may introduce instabilities or require screening mechanisms to satisfy solar system constraints, as discussed by \textit{Clifton et al.}~\cite{Clifton2012} and \textit{Nojiri \& Odintsov}~\cite{NojiriOdintsov2011}.
	
	An alternative geometric formulation of gravity is based on nonmetricity. In symmetric teleparallel gravity, the Levi-Civita connection is replaced by a flat and torsion-free connection, where nonmetricity becomes the only geometric quantity, as introduced by \textit{Nester \& Yo}~\cite{NesterYo1999}. Later, \textit{Beltrán Jiménez, Heisenberg, and Koivisto} developed the coincident formulation of General Relativity~\cite{Jimenez2018} and its cosmological extension in $f(Q)$ gravity~\cite{Jimenez2020}. This framework reveals additional scalar degrees of freedom and provides a consistent description of cosmology in terms of nonmetricity.
	
	More recently, a comprehensive review of $f(Q)$ gravity has been provided by \textit{Heisenberg}~\cite{Heisenberg2024}, while observational applications have been explored by \textit{Kadam, Revanth Kumar, and Yadav}~\cite{Kadam2026}. These developments highlight the growing interest in nonmetricity-based gravity as a viable alternative to curvature-based modifications.

	Moreover, inflationary cosmology describes a very early phase of the Universe during which it underwent an extremely rapid accelerated expansion shortly after the Big Bang. This phenomenon is typically driven by a fundamental scalar field, known as the \textit{inflaton}, which is interpreted as an effective description of vacuum energy.
	
	In this framework, the inflaton behaves as a dynamical scalar field similar to the Higgs field, but governed by a distinct effective potential. During the inflationary epoch, its evolution is characterized by a slow-roll regime, which leads to an effective negative pressure. This property is responsible for a quasi-exponential expansion of spacetime, while the energy density of the field remains nearly constant, mimicking the behavior of a cosmological constant~\cite{Guth1981, Linde1982}.
	
	A key feature of inflation is the generation of primordial perturbations. These arise from quantum fluctuations of the scalar field during the accelerated expansion phase and provide the initial seeds for the formation of large-scale structures in the Universe, including galaxies and the anisotropies of the cosmic microwave background~\cite{Mukhanov1981, Mukhanov1992}.
	
	Observations indicate that these primordial fluctuations are nearly scale-invariant, a crucial requirement for any realistic inflationary model~\cite{Planck2020}. This condition is naturally satisfied in slow-roll inflationary scenarios, where the scalar field evolves gradually along its potential.
	
	Finally, modern cosmological data allow stringent constraints on inflationary models through observable parameters such as the scalar spectral index $n_s$ and the tensor-to-scalar ratio $r$. These observational bounds have led to the development of well-motivated classes of models, including cosmological attractors, which exhibit robust and universal predictions consistent with current data~\cite{PlanckInflationReview}.
	
\noindent
Over the past decades, scalar fields have played a central role in modeling the inflationary phase of the early universe. In particular, within the framework of General Relativity, the standard slow-roll inflation driven by a minimally coupled scalar field has been extensively studied and successfully confronted with observations \cite{Linde1983,LiddleLyth2000}. However, in order to address some theoretical limitations and to explore richer phenomenology, several extensions of gravity have been proposed where the scalar field is nontrivially coupled to geometric quantities.

\medskip

In this context, $f(R)$ gravity has been widely investigated as a natural extension of Einstein's theory. Inflationary scenarios arising from $f(R)$ models, notably the Starobinsky model, have shown remarkable agreement with observational data \cite{Starobinsky1980,DeFeliceTsujikawa2010}. Furthermore, the inclusion of scalar fields in $f(R)$ gravity has led to generalized scalar-tensor theories, where the interplay between curvature and scalar degrees of freedom provides viable mechanisms for both inflation and its graceful exit \cite{NojiriOdintsov2011}.

\medskip

Similarly, modified theories such as $f(T)$ gravity, based on torsion rather than curvature, have also been explored in the context of inflation. In these models, the coupling between a scalar field and the torsion scalar leads to alternative inflationary dynamics, offering new insights into early-universe physics and potentially distinct observational signatures \cite{Cai2016,Bahamonde2017}. These approaches demonstrate that torsional modifications can successfully reproduce inflationary behavior without relying solely on curvature-based frameworks.

\medskip

In addition, the $f(R,T)$ gravity, where $T$ denotes the trace of the energy-momentum tensor, has attracted attention due to its explicit matter-geometry coupling. When combined with scalar fields, such models provide a richer structure that can influence inflationary evolution and the generation of cosmological perturbations \cite{Harko2011,Myrzakulov2015}. These studies highlight the importance of nonminimal couplings in shaping the dynamics of the early universe.

\medskip

Motivated by these developments, it is natural to extend the investigation to other geometric frameworks. In particular, theories based on nonmetricity, such as $f(Q)$ gravity, offer a novel geometric description of gravitation. Therefore, exploring inflationary scenarios in the presence of a scalar field coupled to nonmetricity, namely within $f(Q,\phi)$ gravity, represents a promising direction to further understand the fundamental nature of cosmic inflation.	

\section{Scalar Field Coupled $f(Q)$ Gravity Model}

Modified gravity theories based on nonmetricity have recently attracted significant attention as an alternative geometric description of gravitation. In particular, $f(Q)$ gravity, constructed within the symmetric teleparallel framework, provides a viable extension of General Relativity where gravity is encoded in the nonmetricity scalar $Q$ rather than curvature or torsion \cite{Jimenez2018, Lazkoz2019}.

In order to study early Universe dynamics and inflation, we consider a scalar field $\phi$ nonminimally coupled to the nonmetricity scalar. The total action is given by

\begin{eqnarray}
S = \int d^4x \sqrt{-g} \left[ \frac{1}{2\kappa} f(Q,\phi) + \mathcal{L}_{\phi}  \right],
\label{action_general}
\end{eqnarray}

where $\kappa = 8\pi G$, $\mathcal{L}_m$ denotes the matter Lagrangian, and the scalar field Lagrangian is

\begin{eqnarray}
\mathcal{L}_{\phi} = -\frac{1}{2} \partial_\mu \phi \partial^\mu \phi.
\label{lagrangian_phi}
\end{eqnarray}

We consider the specific functional form

\begin{eqnarray}
f(Q,\phi) = Q + \xi Q \phi^2 + V(\phi),
\label{fQphi}
\end{eqnarray}

where $\xi$ is a dimensionless coupling constant and $V(\phi)$ is the scalar potential.

Thus, the action becomes

\begin{eqnarray}
S = \int d^4x \sqrt{-g} \Bigg[
\frac{1}{2\kappa} \left( Q(1+\xi \phi^2) + V(\phi) \right)
- \frac{1}{2} \partial_\mu \phi \partial^\mu \phi
\Bigg].
\label{action_final}
\end{eqnarray}

The field equations are obtained by varying the action with respect to the metric $g_{\mu\nu}$.

Defining

\begin{eqnarray}
F(\phi) = 1 + \xi \phi^2,
\label{Fphi}
\end{eqnarray}

the variation leads to the modified Einstein equations in $f(Q,\phi)$ gravity:

\begin{eqnarray}
\frac{2}{\sqrt{-g}} \nabla_\alpha \left( \sqrt{-g} F P^{\alpha}_{\ \mu\nu} \right)
+ \frac{1}{2} g_{\mu\nu} \left( Q F + V(\phi) \right)
+ F \left( P_{\mu\alpha\beta} Q_{\nu}^{\ \alpha\beta}
- 2 Q_{\alpha\beta\mu} P^{\alpha\beta}_{\ \ \nu} \right)
\nonumber \\
= - \kappa  T_{\mu\nu}^{(\phi)} ,
\label{Einstein_eq}
\end{eqnarray}

where $P^{\alpha}_{\ \mu\nu}$ is the nonmetricity conjugate defined in symmetric teleparallel gravity \cite{Jimenez2018}.

The energy-momentum tensor of the scalar field is given by

\begin{eqnarray}
T_{\mu\nu}^{(\phi)} =
\partial_\mu \phi \partial_\nu \phi
- g_{\mu\nu} \left[
\frac{1}{2} \partial_\alpha \phi \partial^\alpha \phi
+ V(\phi)
\right].
\label{Tphi}
\end{eqnarray}

Variation of the action with respect to $\phi$ yields

\begin{eqnarray}
\Box \phi - V'(\phi)
+ \frac{\xi}{\kappa} Q \phi = 0,
\label{KleinGordon}
\end{eqnarray}

where $\Box = \nabla_\mu \nabla^\mu$.

Assuming a spatially flat Friedmann–Lemaître–Robertson–Walker (FLRW) metric,

\begin{eqnarray}
ds^2 = -dt^2 + a(t)^2 (dx^2 + dy^2 + dz^2),
\label{FLRW}
\end{eqnarray}

the nonmetricity scalar reduces to

\begin{eqnarray}
Q = -6 H^2,
\label{Q_FLRW}
\end{eqnarray}

where $H = \dot{a}/a$ is the Hubble parameter.

The modified Friedmann equations become

\begin{eqnarray}
3H^2 F &=& \kappa \left(  \frac{1}{2}\dot{\phi}^2 + V(\phi) \right)
- \frac{1}{2} V(\phi),
\label{Friedmann1_explicit}
\\
(2\dot{H} + 3H^2)F &=& -\kappa \left(  \frac{1}{2}\dot{\phi}^2 - V(\phi) \right)
- \frac{1}{2} V(\phi)
- 2\xi H \phi \dot{\phi}.
\label{Friedmann2_explicit}
\end{eqnarray}

The scalar field evolution becomes

\begin{eqnarray}
\ddot{\phi} + 3H\dot{\phi} + V'(\phi)
- \frac{6\xi}{\kappa} H^2 \phi = 0.
\label{KG_FLRW}
\end{eqnarray}

\noindent
\noindent
\noindent
The set of these equations will be used to describe the inflationary era under the slow-roll conditions. By adding Eq.~(\ref{Friedmann1_explicit}) and Eq.~(\ref{Friedmann2_explicit}). This yields:

\begin{eqnarray}
\dot{H} 
&=& -\frac{\kappa}{2F} \dot{\phi}^2 
- \frac{\xi}{F} H \phi \dot{\phi}.
\end{eqnarray}

\noindent
By substituting the explicit expression of $F$, namely $F = 1 + \xi \phi^2$, into the previous equation, we obtain a differential equation governing the evolution of the Hubble parameter, which can be solved once the scalar field dynamics is specified:

\begin{eqnarray}
	2(1 + \xi \phi^2) \, \dot{H} + 2\xi \phi \dot{\phi} \, H + \kappa \dot{\phi}^2 = 0.
	\label{Hdot_standard_form}
\end{eqnarray}
\begin{quote}
	This equation can be tackled in different ways, either by imposing suitable assumptions or by adopting a form of the Hubble parameter consistent with cosmological requirements. In this work, our main objective is to describe the inflationary phase within the framework of $f(Q)$ gravity. However, it remains challenging in the literature to propose a general expression for $\phi(t)$ that naturally accounts for this period. On the other hand, several studies have relied on specific choices of the scale factor — or equivalently the expansion rate — capable of reproducing an inflationary dynamics, notably the de Sitter scenario and power-law inflation models.
\end{quote}
\section{de Sitter Inflation}

The de Sitter regime plays a fundamental role in cosmology, as it provides an excellent description of the rapid exponential expansion that is believed to have occurred in the early Universe (\cite{Haba2015}). This phase is commonly modeled through a scale factor of exponential form given by
\begin{eqnarray}
a(t) = A \, e^{H_0 t},
\end{eqnarray}
where $A$ and $H_0$ are positive constants. Such a choice immediately implies a constant expansion rate, since the Hubble parameter becomes
\begin{eqnarray}
H(t) = \frac{\dot{a}}{a} = H_0.
\label{Sitter_H}
\end{eqnarray}

By substituting this constant Hubble parameter into the differential equation~(\ref{Hdot_standard_form}), one can determine the corresponding evolution of the scalar field during the inflationary period.

\begin{eqnarray}
\phi(t) = \phi_0 \, \exp\left(- \frac{2\xi H_0}{\kappa} t \right).
\label{phi_solution_desitter}
\end{eqnarray}
	
where $\phi_0$ is an integration constant. Under the de Sitter evolution, where the Hubble parameter is constant as in Eq.~(\ref{Sitter_H}), the solution of the Klein--Gordon equation (\ref{KG_FLRW}) leads to the scalar field potential.	
	
\begin{eqnarray}
V(t) &=& \frac{1}{2} \left( \frac{12 \, \xi \, H_0^2}{\kappa} - \frac{4 \, \xi^2 \, H_0^2}{\kappa^2} \right) \phi_0^2 \, \exp\left(- \frac{4 \, \xi \, H_0}{\kappa} t \right) + C_{1},
\label{V_desitter}
\end{eqnarray}
where $C_{1}$ is an integration constant.

\noindent
The study of inflation in cosmology allows one to confront theoretical predictions with observational evidence. It is widely discussed in the literature~\cite{Linde2014} that in standard scalar field models of inflation, key inflationary observables—such as the spectral index, the tensor-to-scalar ratio, and the running of the spectral index—can be expressed in terms of the scalar field potential $V(\phi)$. Moreover, depending on the slow-roll parameters, these quantities may also be written in terms of the Hubble parameter. Initially, it is necessary to define the slow-roll parameters in relation to the canonical scalar field potential $V(\phi)$ as follows:
\begin{eqnarray}
\epsilon &=& \frac{1}{2} \left( \frac{V'(\phi)}{V(\phi)} \right)^2, \\
\eta &=& \frac{1}{2} \frac{V''(\phi)}{V(\phi)}.
\label{eq:slow_roll_parameters}
\end{eqnarray}

\noindent
Based on the scalar field potential given in (\ref{V_desitter}), the slow-roll parameters are expressed as
\noindent

\begin{eqnarray}
\epsilon(\phi) &=& \frac{1}{2} \left( \frac{m_\text{eff}^2 \phi}{\frac{1}{2} m_\text{eff}^2 \phi^2 + C_{1}} \right)^2 
= \frac{2 \phi^2}{( \phi^2 + 2 C_{1} / m_\text{eff}^2 )^2}, \\
\eta(\phi) &=& \frac{m_\text{eff}^2}{\frac{1}{2} m_\text{eff}^2 \phi^2 + C_{1}} 
= \frac{2}{\phi^2 + 2 C_{1} / m_\text{eff}^2}.
\end{eqnarray}
where $m_\text{eff}^2 = 3 H_0 (\kappa + 6 \xi) - (\kappa + 6 \xi)^2 + \frac{\kappa}{6 \xi} H_0^2$.\\
\noindent
From these, the inflationary observables are derived as
\begin{eqnarray}
n_s &=& 1 - 6 \epsilon + 2 \eta, \\
r &=& 16 \epsilon.
\end{eqnarray}

\noindent
These expressions provide a direct link between the scalar field dynamics in the modified $f(Q)$ gravity model and the observable quantities constrained by cosmological data, allowing us to test the viability of the model against Planck or other CMB observations.

According to (\ref{V_desitter}), they give
\begin{eqnarray}
n_s(\phi) &=& 1 + \frac{8\left(\frac{C_1}{m_{\text{eff}}^2} - \phi^2\right)}
{\left(\phi^2+\frac{2C_1}{m_{\text{eff}}^2}\right)^2}
\label{ns_final}
\end{eqnarray}

\begin{eqnarray}
r(\phi) &=& \frac{32 \phi^2}
{\left(\phi^2+\frac{2C_1}{m_{\text{eff}}^2}\right)^2}
\label{r_final}
\end{eqnarray}

Following a procedure similar to that adopted in \cite{Odintsov2023}, one can rewrite the inflationary observables as functions of the number of e-folds $N$. This parametrization is particularly useful to study their evolution and to evaluate them at the end of inflation in the context of cosmological models arising from modified  gravity.

Starting from Eq.~(\ref{Sitter_H}), the number of e-folds can be expressed either as a function of cosmic time or directly in terms of the scalar field. One has
\begin{eqnarray}
N &=& \int_{t_i}^{t_f} H(t)\, dt = \int_{t_i}^{t_f} H_0\, dt = H_0 (t_f - t_i),
\label{N_time}
\end{eqnarray}

Using the scalar field evolution given in Eq.~(\ref{phi_solution_desitter}), we can rewrite $N$ in terms of $\phi$ as
\begin{eqnarray}
N &=& \int_{\phi_i}^{\phi_f} \frac{H}{\dot{\phi}}\, d\phi 
= -\frac{1}{\alpha} \int_{\phi_i}^{\phi_f} \frac{d\phi}{\phi}
= \frac{\kappa}{2\xi} \ln\left(\frac{\phi_i}{\phi_f}\right),
\label{N_phi}
\end{eqnarray}
where $\phi_i$ and $\phi_f$ denote respectively the values of the scalar field at the beginning and at the end of inflation.

In the standard inflationary framework, the end of inflation is defined by the condition that the first slow-roll parameter becomes of order unity, namely $\epsilon(\phi_f) \simeq 1$. Using the expression of $\epsilon(\phi)$ obtained previously, this condition leads to
\begin{eqnarray}
\phi_f^2 = (1 - A) \pm \sqrt{1 - 2A}.
\label{phi_f_final}
\end{eqnarray}

with
\begin{eqnarray}
A = \frac{2C_1}{m_{\text{eff}}^2},
\label{A_def}
\end{eqnarray}

from (\ref{N_phi}) the scalar field can be expressed as a function of $N$ as
\begin{eqnarray}
\phi(N) = \phi_f \exp\left(\frac{2\xi}{\kappa} N\right).
\label{phi_N_expression}
\end{eqnarray}

Using the expression of the scalar field given in Eq.~(\ref{phi_N_expression}), we can now express the inflationary observables as functions of the number of e-folds $N$. Substituting $\phi(N)$ into the slow-roll parameters, we first obtain
\begin{eqnarray}
\phi^2(N) = \phi_f^2 \exp\left(\frac{4\xi}{\kappa} N\right).
\label{phi2_N}
\end{eqnarray}

Therefore, the scalar spectral index $n_s$ can be written as
\begin{eqnarray}
n_s(N) &=& 1 + \frac{8\left(\frac{C_1}{m_{\text{eff}}^2} - \phi_f^2 e^{\frac{4\xi}{\kappa} N}\right)}
{\left(\phi_f^2 e^{\frac{4\xi}{\kappa} N}+\frac{2C_1}{m_{\text{eff}}^2}\right)^2}.
\label{ns_N_final}
\end{eqnarray}

Similarly, the tensor-to-scalar ratio $r$ takes the form
\begin{eqnarray}
r(N) &=& \frac{32 \, \phi_f^2 e^{\frac{4\xi}{\kappa} N}}
{\left(\phi_f^2 e^{\frac{4\xi}{\kappa} N}+\frac{2C_1}{m_{\text{eff}}^2}\right)^2}.
\label{r_N_final}
\end{eqnarray}
We analyze the effect of the nonminimal coupling parameter $\xi$ on inflationary observables. In the regime $\xi \ll 1$, we have $m_\text{eff}^2 \approx 6 \xi H_0^2 / \kappa$, leading to $A(\xi) \approx 12 C_1 \xi H_0^2 / \kappa$, hence $A \propto \xi$. For small $\xi$, the correction to $\phi^2(N)$ is negligible, yielding a quasi-constant regime with $r \sim \mathcal{O}(0.01)$ and $n_s \lesssim 1$, compatible with Planck observations. Conversely, for large $\xi$, $m_\text{eff}^2 \approx -36 \xi^2$ and $A \approx -18 C_1 \xi^2$, making $A$ negative and very small; here, $\beta \gg 1$ ($\beta = \frac{4 \, \xi}{\kappa}$) and $\phi^2(N) \gg 1$, resulting in $r \to 0$ and $n_s \to 1$, which is only partially compatible with data. The intermediate regime, considered optimal, corresponds to $\xi$ values such that $\beta N \sim \mathcal{O}(1)$ for $N=60$, i.e., $\xi \sim 10^{-3}\kappa - 10^{-2}\kappa$. In this interval, $e^{\beta N} \sim 2-10$, $A(\xi)$ remains positive and below $1/2$, and $\phi^2(N) \sim A$, giving typical values $n_s \approx 0.96-0.968$ and $r \approx 0.01-0.05$, making the model compatible with Planck and BICEP constraints for intermediate $\xi$.

In conclusion, the analysis shows that the model naturally selects a narrow range of the nonminimal coupling parameter $\xi$ that satisfies the current constraints on the inflationary observables. Values of $\xi$ outside this range either produce a scalar spectral index $n_s > 1$ (blue spectrum) or a tensor-to-scalar ratio $r$ too large, both in conflict with Planck/BICEP results. Therefore, the inflationary scenario in the present $f(Q)$ framework is observationally viable only for
\begin{equation}
10^{-3} \kappa \lesssim \xi \lesssim 10^{-2} \kappa.
\end{equation}

This result highlights the predictive power of the nonminimal derivative coupling and its critical role in shaping the inflationary dynamics within modified $f(Q)$ gravity.

\clearpage
\begin{figure}[htbp]
	\centering
	
	\begin{subfigure}{0.6\textwidth}
		\centering
		\includegraphics[width=\linewidth]{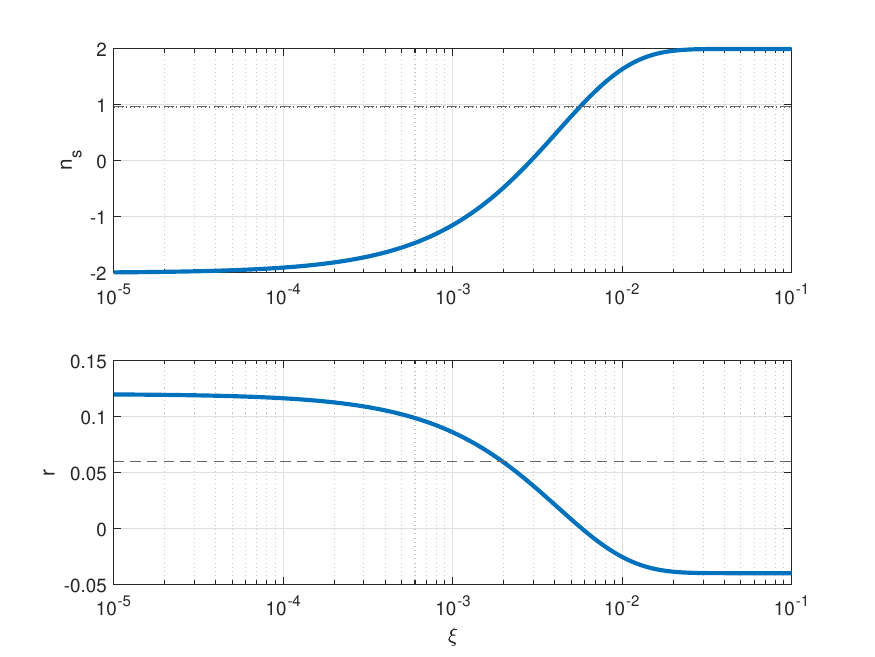}
	\end{subfigure}
\caption{Numerical evolution of the inflationary observables. 
	Top panel: scalar spectral index $n_s$ as a function of $\xi$. 
	Bottom panel: tensor-to-scalar ratio $r$ as a function of $\xi$. 
	The parameters are fixed to $\kappa = 1$, $C_1 = 10^{-6}$, $H_0 = 10^{-5}$, and $N = 60$.}
	\label{figure_1_deSitter}
\end{figure}

\begin{figure}[htbp]
	\hfill
		\centering
		\includegraphics[width=\linewidth]{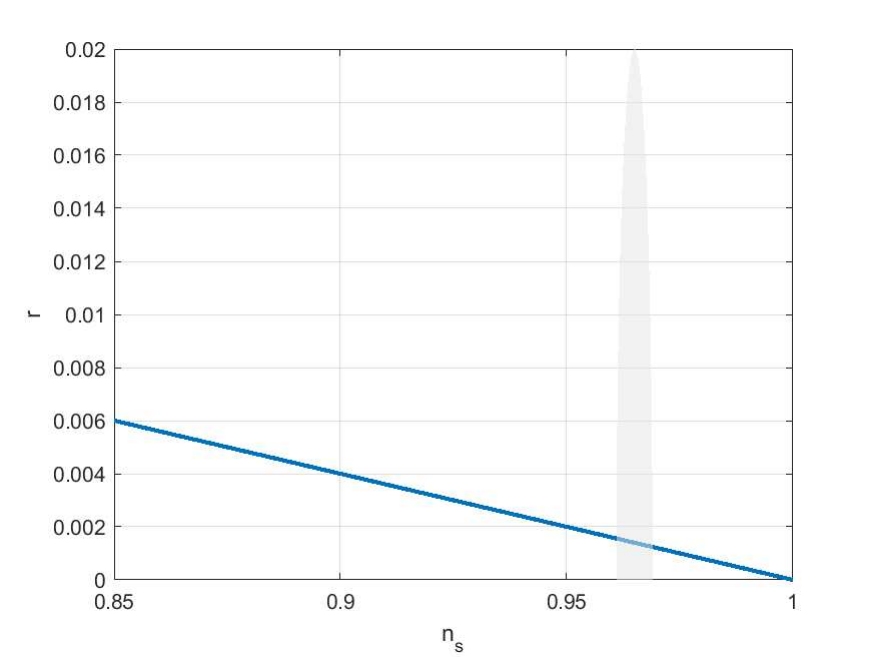}
		\caption{
			Tensor-to-scalar ratio $r$ versus scalar spectral index $n_s$ for $\kappa = 1$, $C_1 = 10^{-6}$, and $H_0 = 10^{-5}$.
		}
	
	\label{figure_2_deSitter}
\end{figure}

\clearpage

Within the framework of $f(Q,\phi)$ gravity, we now consider the case of \textit{De Sitter inflation} and investigate the impact of the nonminimal coupling parameter $\xi$ on the inflationary observables $n_s$ and $r$. The results are presented in Fig \ref{figure_1_deSitter}, \ref{figure_2_deSitter}.

The upper panel shows the evolution of the scalar spectral index $n_s$ as a function of $\xi$ (in logarithmic scale). One observes that $n_s$ increases monotonically with $\xi$, starting from negative values for very small $\xi \sim 10^{-5}$ and crossing the scale-invariant limit $n_s = 1$ around $\xi \sim 10^{-3}$. For larger values of $\xi$, the model predicts $n_s > 1$, corresponding to a blue-tilted spectrum, which is disfavored by current cosmological observations.

The middle panel displays the behavior of the tensor-to-scalar ratio $r(\xi)$. In contrast to $n_s$, the quantity $r$ decreases as $\xi$ increases. It starts from relatively large values $r \sim 0.12$ at $\xi \sim 10^{-5}$ and gradually decreases, crossing zero and becoming slightly negative for $\xi \gtrsim 10^{-2}$, which is unphysical. This indicates that large values of $\xi$ are not viable within this framework.

The parametric plot $r(n_s)$ shown in the lower panel highlights a clear inverse correlation between the two observables. As $n_s$ approaches the observationally favored region $n_s \approx 0.965$, the tensor-to-scalar ratio is strongly suppressed, reaching values of the order
\begin{equation}
r \sim 10^{-3} - 10^{-2}.
\end{equation}

The shaded region corresponds to the Planck observational constraints, indicating that only a narrow range of $\xi$ is compatible with the data. More precisely, the viable region is approximately given by
\begin{equation}
10^{-3} \lesssim \xi \lesssim 10^{-2},
\end{equation}
for which the model predicts
\begin{equation}
n_s \approx 0.96 - 0.97, \qquad r \lesssim 0.01.
\end{equation}

These results show that, although the De Sitter inflationary scenario in $f(Q,\phi)$ gravity can reproduce observationally consistent values of $n_s$ and $r$, this agreement is restricted to an intermediate regime of the coupling parameter $\xi$. For very small $\xi$, the tensor modes are too large, while for large $\xi$, the scalar spectrum becomes blue-tilted and the tensor-to-scalar ratio loses its physical meaning.

In conclusion, the De Sitter case within $f(Q,\phi)$ gravity remains viable only in a constrained parameter space, highlighting again the crucial role played by the nonminimal coupling in shaping the inflationary dynamics and ensuring compatibility with observations.

\section{Power-Law Inflation}
Power-law inflation~\cite{Goheer2009} is characterized by a scale factor evolving as a power of cosmic time, namely
\begin{eqnarray}
a(t) = A\, t^{p}, \qquad H(t) = \frac{p}{t}.
\label{power_law_background}
\end{eqnarray}

Adopting this framework, we proceed in a manner analogous to the de Sitter case. In particular, substituting this ansatz into the dynamical equation~(\ref{Hdot_standard_form}) leads to the corresponding differential equation governing the evolution of the scalar field.

We consider the scalar field equation with non-minimal derivative coupling (\ref{Hdot_standard_form}) and assume power-law inflation (\ref{power_law_background}) we obtain
\begin{equation}
\kappa \dot{\phi}^2 + \frac{2p \xi}{t} \phi \dot{\phi} - \frac{2p}{t^2}(1+\xi \phi^2) = 0,
\label{eq:scalarevolution}
\end{equation}

\noindent
The scalar field evolution equation~\eqref{eq:scalarevolution} is a nonlinear differential equation due to the presence of both $\dot{\phi}^2$ and the term $\phi \dot{\phi}$ coupled with $\xi$. Its exact analytical solution is therefore not straightforward and cannot be expressed in a simple closed form. To proceed, we consider a \textbf{physically motivated approximation} valid in the regime of small non-minimal coupling, $\xi \phi^2 \ll 1$. Under this assumption, the term $1 + \xi \phi^2$ can be approximated by unity, which significantly simplifies the equation and allows us to derive an analytical expression for $\phi(t)$. This approach preserves the main physical features while keeping the analysis tractable for further calculations, such as reconstructing the potential $V(\phi)$.

In the weak coupling regime $\xi \phi^2 \ll 1$, we approximate $1 + \xi \phi^2 \simeq 1$, which simplifies the equation \ref{eq:scalarevolution} to
\begin{equation}
t \frac{d\phi}{dt} + \frac{p\xi}{\kappa} \phi = \pm \sqrt{\frac{2p}{\kappa}}.
\end{equation}

This is a linear differential equation with solution
\begin{equation}
\phi(t) \approx \pm \sqrt{\frac{2p}{\kappa}} \, t + C_{2} \, t^{-p\xi/\kappa},
\label{eq:phi_powerlaw_solution}
\end{equation}
where $C_{2}$ is an integration constant.

It will be interesting to complete our analysis in this section by reconstructing the $f(Q)$ models that support power-law inflation. We recall that the scalar field can be directly related to the first derivative of the function $f(Q)$. Therefore, it is sufficient to express the scalar field as a function of the non-metricity scalar $Q$. In the present context of power-law expansion, we use the expression of the scalar field given in (\ref{eq:phi_powerlaw_solution}) along with the Hubble parameter from (\ref{power_law_background}). From (\ref{power_law_background}) and (\ref{Q_FLRW}), we obtened 
 \begin{equation}
 t = p \, 6^{1/2} (-Q)^{-1/2},
 \label{eq:time_vs_Q}
 \end{equation}
By substituting Eq.~(\ref{eq:time_vs_Q}) into Eq.~(\ref{eq:phi_powerlaw_solution}), the scalar field as a function of the non-metricity scalar $Q$ is obtained as follows:
\begin{equation}
\phi(Q) \approx \pm \frac{\sqrt{12}\, p \sqrt{p}}{\sqrt{\kappa}} \, (-Q)^{-1/2} 
+ C_2 \, (\sqrt{6}\, p)^{-p\xi/\kappa} \, (-Q)^{p\xi/(2\kappa)},
\label{eq:phi_vs_Q}
\end{equation}

\noindent
In the regime where the second term dominates, the scalar field expression in Eq.~(\ref{eq:phi_vs_Q}) reduces to

Using the full expression of the scalar field $\phi(Q)$, the function $f(Q)$ is obtained by integrating $F(\phi)=\frac{df}{dQ}=1+\xi \phi^2$. This leads to the general expression
Using the full expression of the scalar field, the function $f(Q)$ can be written explicitly as
\begin{equation}
\boxed{\begin{aligned}
f(Q) =\;& Q - \frac{12 \xi p^3}{\kappa} \ln(-Q) \\
&+ \frac{4 \kappa \xi C_2}{p\xi + \kappa} 
\frac{\sqrt{12}\, p \sqrt{p}}{\sqrt{\kappa}} 
(\sqrt{6}\, p)^{-p\xi/\kappa} (-Q)^{\frac{p\xi}{2\kappa} + \frac{1}{2}}
+ \frac{\xi C_2^2}{p\xi + \kappa} 
(\sqrt{6}\, p)^{-2p\xi/\kappa} (-Q)^{\frac{p\xi}{\kappa} + 1}
+ C_{3}.
\end{aligned}}
\label{eq:fQ_compact}
\end{equation}

Using the approximate solution
\begin{equation}
\phi(t) \approx A t, \quad \text{with} \quad A = \pm \sqrt{\frac{2p}{\kappa}},
\label{eq:attractor_solution}
\end{equation}

\noindent

This approximation is justified as follows. From the general solution $\phi(t) = A t + C_{2} \, t^{-p\xi/\kappa}$, the second term corresponds to a decaying mode provided that $p\xi/\kappa > 0$. During the inflationary regime (large cosmic time), this contribution rapidly becomes negligible compared to the linear term, so that the scalar field dynamics is effectively governed by the attractor solution $\phi(t) \sim A t$. This behavior is consistent with the slow-roll regime, where the scalar field evolves smoothly and subleading corrections decay with time. Such attractor solutions are widely used in inflationary cosmology to derive analytical results (see, e.g., \cite{LiddleLyth2000, Copeland1998, Tsujikawa2013}). Therefore, retaining only the leading term provides a physically well-motivated and reliable approximation for reconstructing the potential $V(\phi)$.\\

Together with the Klein--Gordon equation (\ref{KleinGordon}),equation (\ref{eq:attractor_solution}) and $H = \frac{p}{t}$, we reconstruct the scalar potential takes the form
\begin{equation}
V(\phi) = \frac{2p}{\kappa}
\left(-3p + \frac{6\xi p^2}{\kappa} \right)\ln \phi + V_0,
\label{eq:logarithmic_potential}
\end{equation}
where $V_0$ is an integration constant.

By combining Eqs.~\eqref{eq:slow_roll_parameters} and \eqref{eq:logarithmic_potential}, the slow-roll parameters can be directly expressed as functions of the scalar field $\phi$, as shown above.
\begin{eqnarray}
\epsilon(\phi) &=& \frac{1}{2}
\frac{\left[\frac{2p}{\kappa}\left(-3p + \frac{6\xi p^2}{\kappa}\right)\right]^2}
{\phi^2 \left[\frac{2p}{\kappa}\left(-3p + \frac{6\xi p^2}{\kappa}\right)\ln\phi + V_0\right]^2}
\label{eq:epsilon_log_potential}
\end{eqnarray}
\begin{eqnarray}
\eta(\phi) &=& -\frac{1}{2}
\frac{\frac{2p}{\kappa}\left(-3p + \frac{6\xi p^2}{\kappa}\right)}
{\phi^2 \left[\frac{2p}{\kappa}\left(-3p + \frac{6\xi p^2}{\kappa}\right)\ln\phi + V_0\right]}.
\label{eq:explicit_slow_roll_log_potential}
\end{eqnarray}

Using the definition of the number of e-folds together with the dominant approximation of the scalar field, one directly obtains an expression of the number of e-folds as a function of the scalar field $\phi$, given by
\begin{equation}
N = p \ln\left(\frac{\phi_f}{\phi_i}\right).
\label{eq:efolds_field_relation}
\end{equation}

In the slow-roll regime, the first slow-roll parameter \ref{eq:epsilon_log_potential} reduces to
\begin{equation}
\epsilon(\phi) = \frac{1}{2}\frac{C^2}{\phi^2 \left(C\ln\phi + V_0\right)^2}.
\end{equation}

with
\begin{equation}
C = \frac{2p}{\kappa}\left(-3p + \frac{6\xi p^2}{\kappa}\right),
\end{equation}

At the end of inflation, the condition
$\epsilon(\phi_f)=1$ $\Longrightarrow \phi = \phi_f$. Assuming the slow-roll condition such that the constant term dominates,
\begin{equation}
|C \ln\phi| \ll V_0,
\end{equation}
 gives
\begin{equation}
\phi_f \approx \frac{|C|}{\sqrt{2}\,V_0}.
\end{equation}

Substituting $\phi_f$ in (\ref{eq:efolds_field_relation}) the final expression reads
\begin{equation}
\phi_i(N) =
\frac{1}{\sqrt{2}\,V_0}
\left|
\frac{2p}{\kappa}\left(-3p + \frac{6\xi p^2}{\kappa}\right)
\right|
\, e^{-\frac{N}{p}}.
\label{eq:phi_initial}
\end{equation}

In the slow-roll regime, by substituting the initial field value \(\phi_i(N)\) from Eq.~\eqref{eq:phi_initial} into the explicit slow-roll parameters, one obtains the spectral index and tensor-to-scalar ratio as functions of the number of e-folds \(N\):

\begin{equation}
r(N) = \frac{16 \, V_0^2 \, e^{\frac{2N}{p}}}
{\left[ B \ln\left(\frac{|B|}{\sqrt{2}\, V_0}\right) - \frac{B N}{p} + V_0 \right]^2}
\label{eq:r_N}
\end{equation}

\begin{equation}
n_s(N) = 1 - \frac{2 V_0^2 \, e^{\frac{2N}{p}} \left( 4 B + V_0 - \frac{B N}{p} + B \ln \frac{|B|}{\sqrt{2} V_0} \right)}
{B \left( B \ln \frac{|B|}{\sqrt{2} V_0} - \frac{B N}{p} + V_0 \right)^2}
\label{eq:ns_N_compact}
\end{equation}

with 
\begin{equation}
B = \frac{2p}{\kappa}\left(-3p + \frac{6\xi p^2}{\kappa}\right).
\end{equation}

\clearpage
\begin{figure}[htbp]
	\centering
	
	\begin{subfigure}{0.48\textwidth}
		\centering
		\includegraphics[width=\linewidth]{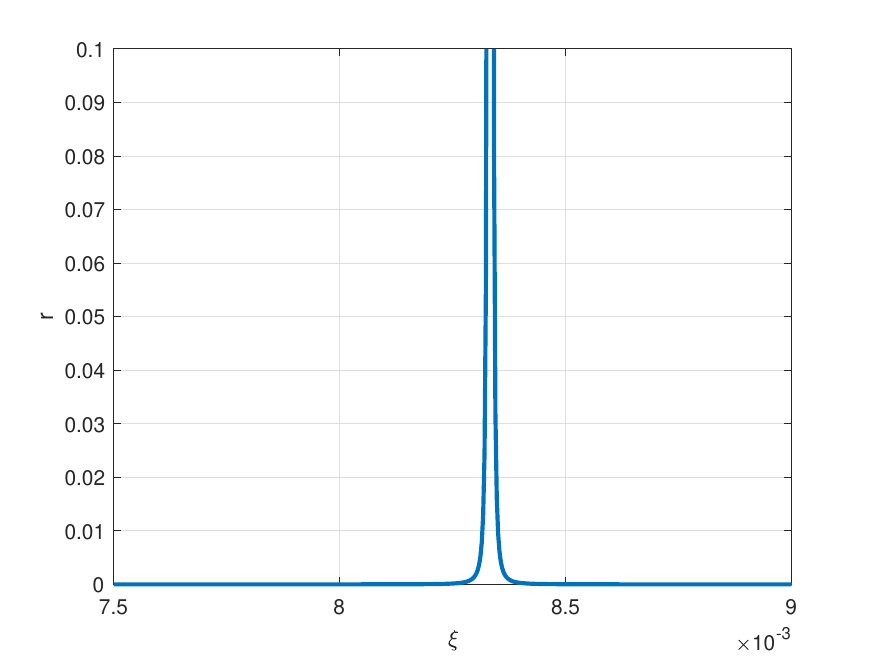}
	\end{subfigure}
	\hfill
	\begin{subfigure}{0.48\textwidth}
		\centering
		\includegraphics[width=\linewidth]{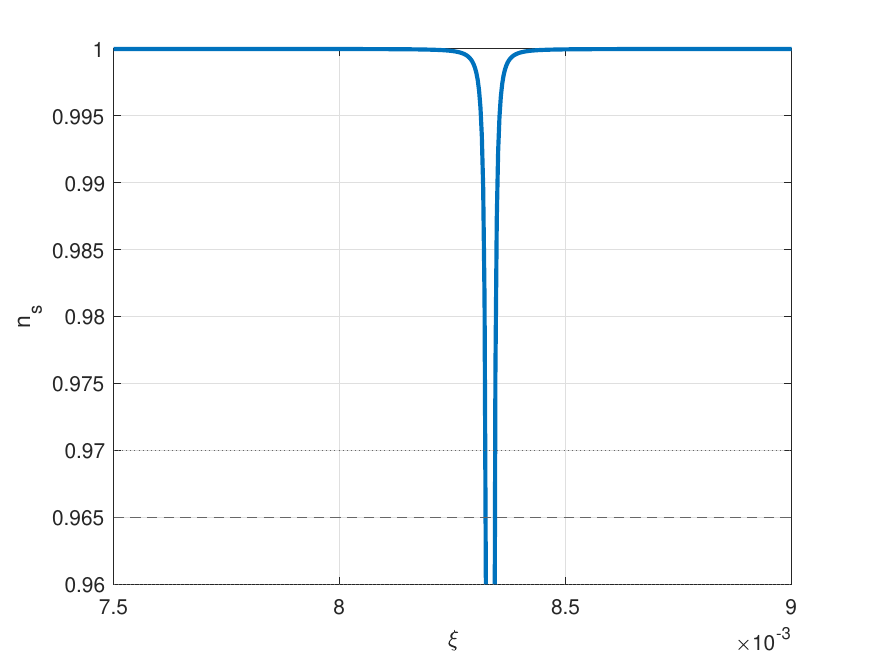}
	\end{subfigure}
	
	\caption{Numerical evolution of the inflationary observables. Left panel: tensor-to-scalar ratio $r$ as a function of $\xi$. Right panel: scalar spectral index $n_s$ as a function of $\xi$. The parameters are fixed to $\kappa = 1$, $p = 60$, and $V_0 = 1$.}
	\label{figure_1_pawerlow}
\end{figure}

\begin{figure}[htbp]
	\centering
	\includegraphics[width=0.7\textwidth]{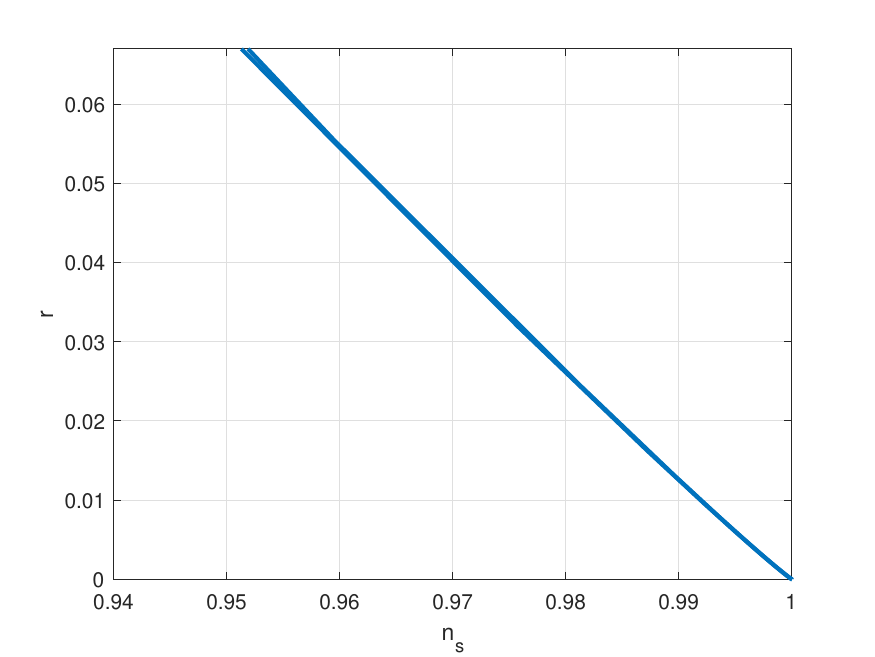}
	\caption{
			Tensor-to-scalar ratio $r$ as a function of the scalar spectral index $n_s$ for $\kappa = 1$, $p = 60$, and $V_0 = 1$.
		}
	\label{figure_2_pawerlow}
\end{figure}

\clearpage

In order to ensure the viability of the model with respect to observational data, it is necessary to constrain the coupling parameter $\xi$ using the predictions for the scalar spectral index $n_s$ and the tensor-to-scalar ratio $r$. From the definition of the parameter $B$, one finds that its sign and magnitude play a crucial role in determining the behavior of both $n_s$ and $r$, and a physically consistent inflationary regime requires $B$ to be negative and sufficiently small in magnitude. This condition translates into a constraint on the coupling parameter $\xi$, leading to $\xi < \frac{\kappa}{2p}$. For the specific choice $\kappa = 1$ and $p = 60$, this upper bound becomes $\xi < \frac{1}{120} \approx 0.00833$. Furthermore, compatibility with observational constraints from Planck and BICEP/Keck data, which favor $n_s \simeq 0.965$ and a small tensor-to-scalar ratio $r < 0.036$, requires $\xi$ to lie close to this upper bound. As a result, the allowed range of the coupling parameter can be expressed as $0 < \xi < 0.00833$, with a preferred region given approximately by $\xi \sim \mathcal{O}(10^{-3})$. It is worth noting that a similar choice of parameters, namely $\kappa = 1$ and $p = 60$, has been adopted in the work of S.~Haba \textit{et al.} \cite{Haba2025}, where these values were shown to provide a good agreement with Planck observational constraints, further supporting the consistency of the present analysis.

For $N = 60$, the inflationary observables strongly depend on the coupling parameter $\xi$. In particular, near the critical value $\xi_c = 1/(2p)$, the model exhibits a sensitive behavior. Away from this point, one typically obtains
\begin{equation}
n_s \approx 0.96 - 0.97, \quad r \lesssim 10^{-2},
\end{equation}
which are consistent with current observational constraints.

\section{Cosh-Type Inflation}

Assuming the scale factor
\begin{eqnarray}
a(t) = \gamma \cosh(\gamma t),
\label{scale_factor_cosh}
\end{eqnarray}
the Hubble parameter and its time derivative are given by
\begin{eqnarray}
H(t) = \gamma \tanh(\gamma t).
\label{Hubble_and_Hdot_cosh}
\end{eqnarray}

Substituting into Eq.~(\ref{Hdot_standard_form})

\begin{eqnarray}
(1 + \xi \phi^2)\gamma^2 \mathrm{sech}^2(\gamma t)
+ \xi \gamma \tanh(\gamma t)\, \phi \dot{\phi}
+ \frac{\kappa}{2} \dot{\phi}^2 = 0.
\label{phi_differential_eq}
\end{eqnarray}

 \begin{equation}
 \text{In the asymptotic regime } t \gg 1/\gamma,\; \tanh(\gamma t) \to 1 \text{ and } \mathrm{sech}^2(\gamma t) \to 0,\; \text{which simplifies the equation to}
 \end{equation} the scalar field evolution reduces to
\begin{eqnarray}
\dot{\phi} = - \frac{2\xi \gamma}{\kappa}\phi,
\label{phi_dot_final}
\end{eqnarray}
which yields the solution
\begin{eqnarray}
\phi(t) = \phi_0 \exp\left(-\frac{2\xi \gamma}{\kappa} t\right).
\label{phi_final}
\end{eqnarray}

Using the slow-roll condition applied to Eq.~(\ref{KG_FLRW}), together with the previous expression of $\phi$, the approximation $t \gg 1/\gamma$ (where $\tanh(\gamma t) \to 1$), and Eq.~(\ref{Hubble_and_Hdot_cosh}), the potential is obtained as
\begin{eqnarray}
V(\phi) = \frac{6\xi \gamma^2}{\kappa} \phi^2 + C_{4}.
\label{V_final}
\end{eqnarray}
\text{where $C_{4}$ is the integration constant.}

In the following, we aim to reconstruct the function $f(Q)$ in the context of the non-minimal derivative coupling model. 
To do so, we first express the cosmic time $t$ in terms of the nonmetricity scalar $Q$. Starting from the FLRW relation (\ref{Q_FLRW}) and using the Hubble parameter  
\ref{Hubble_and_Hdot_cosh}, we can invert the relation to write $t$ as a function of $H$ and subsequently as a function of $Q$. This gives
\begin{equation}
t = \frac{1}{\gamma} \, \text{arctanh}\left(\frac{H}{\gamma}\right) 
= \frac{1}{\gamma} \, \text{arctanh}\left(\frac{\sqrt{-Q/6}}{\gamma}\right).
\label{t_of_Q_tanh}
\end{equation}
This expression will serve as the basis to rewrite all relevant quantities, such as the scalar field and potential, in terms of $Q$, facilitating the reconstruction of $f(Q)$.

Starting from the solution of the scalar field 
\begin{equation}
\phi(t) = \phi_0 \exp\left(-\frac{2\xi \gamma}{\kappa} t\right),
\label{phi_final}
\end{equation}
and substituting the expression of $t$ as a function of $Q$ from (\ref{t_of_Q_tanh}), we obtain
\begin{equation}
\phi(Q) = \phi_0 \exp\left[-\frac{2 \xi \gamma}{\kappa} \frac{1}{\gamma} \, \text{arctanh}\left(\frac{\sqrt{-Q/6}}{\gamma}\right)\right]
= \phi_0 \exp\left[-\frac{2 \xi}{\kappa} \, \text{arctanh}\left(\frac{\sqrt{-Q/6}}{\gamma}\right)\right].
\label{phi_Q}
\end{equation}

In this context, we aim to reconstruct the function $f(Q)$ within the framework of a non-minimal coupling between the scalar field and the non-metricity scalar. Following the approach $F(\phi) = 1 + \xi \phi^2$, it is natural to consider the ansatz
\begin{equation}
f(Q) = F(\phi) \, Q + C_{5},
\end{equation}
where $C_{5}$ is an integration constant. Using the expression of $\phi$ in terms of $Q$, obtained from (\ref{Q_FLRW}) and the Hubble solution (\ref{Hubble_and_Hdot_cosh}),
\begin{equation}
\phi(Q) = \phi_0 \exp\left[-\frac{2 \xi}{\kappa} \, \text{arctanh}\left(\frac{\sqrt{-Q/6}}{\gamma}\right)\right],
\end{equation}
we immediately obtain the reconstructed form of $f(Q)$ as
\begin{equation}
\boxed{
	f(Q) = \left[ 1 + \xi \, \phi_0^2 \exp\left(-\frac{4 \xi}{\kappa} \, \text{arctanh}\left(\frac{\sqrt{-Q/6}}{\gamma}\right)\right) \right] Q + C_{5} }.
\end{equation}
This expression explicitly shows how the non-minimal coupling of the scalar field influences the functional form of $f(Q)$.

Using the potential given in Eq.~(\ref{V_final}) together with the definitions of the slow-roll parameters in Eq.~(\ref{eq:slow_roll_parameters}), we obtain the explicit expressions of the inflationary parameters. Therefore, the slow-roll parameters $\epsilon$ and $\eta$ can be written as

\begin{equation}
\epsilon = \frac{72 \xi^2 \gamma^4 \phi^2}{\left(6\xi \gamma^2 \phi^2 + \kappa C_4\right)^2},
\label{eq:epsilon_final}
\end{equation}

\begin{equation}
\eta = \frac{6\xi \gamma^2}{6\xi \gamma^2 \phi^2 + \kappa C_4}.
\label{eq:eta_final}
\end{equation}

Using the expressions of (\ref{eq:epsilon_final}) and (\ref{eq:eta_final}) , the inflationary observables can be written explicitly as

\begin{equation}
n_s = 1
- \frac{432 \xi^2 \gamma^4 \phi^2}
{\kappa^2 \left(6\xi \gamma^2 \phi^2 + \kappa C_4\right)^2}
+ \frac{12\xi \gamma^2}
{6\xi \gamma^2 \phi^2 + \kappa C_4},
\label{eq:ns_final_compact}
\end{equation}

\begin{equation}
r = \frac{1152 \xi^2 \gamma^4 \phi^2}
{\kappa^2 \left(6\xi \gamma^2 \phi^2 + \kappa C_4\right)^2}.
\label{eq:r_final_compact}
\end{equation}
The number of e-folds $N$ is obtained from the scalar field dynamics and is given by the following expression:
\begin{equation}
N = \frac{\phi_\text{ini}^2 - \phi_\text{end}^2}{4} + \frac{\kappa C_4}{12 \xi \gamma^2} \ln\left(\frac{\phi_\text{ini}}{\phi_\text{end}}\right)
\label{N_efolds}
\end{equation}

Neglecting the final value of the scalar field $\phi_f$ with respect to the initial value $\phi_i$, the number of e-folds derived from (\ref{V_final}) simplifies to
\begin{equation}
N \approx \frac{\phi_i^2}{4} + \frac{\kappa C_4}{12 \xi \gamma^2} \ln(\phi_i).
\end{equation}
This relation defines a transcendental equation for $\phi_i$. By introducing the change of variable $x = \phi_i^2$, the equation can be rewritten in a suitable form for the application of the Lambert $W$ function. Solving analytically, we obtain the initial scalar field as a function of the number of e-folds:
\begin{equation}
\phi_i(N) = \sqrt{ \frac{\kappa C_4}{6 \xi \gamma^2} \,
	W\left( \frac{6 \xi \gamma^2}{\kappa C_4} \,
	\exp\left(\frac{24 \xi \gamma^2}{\kappa C_4} N\right) \right) },
\label{phi_ini_Lambert}
\end{equation}
where $W(x)$ denotes the Lambert $W$ function.

\clearpage
\begin{figure}[htbp]
	\centering
	
	\begin{subfigure}{0.48\textwidth}
		\centering
		\includegraphics[width=\linewidth]{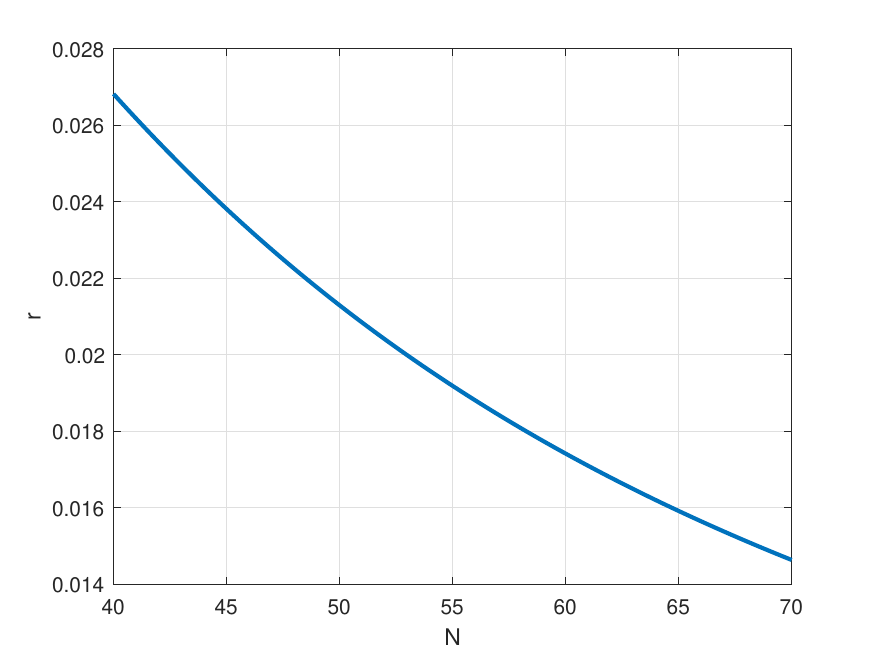}
	\end{subfigure}
	\hfill
	\begin{subfigure}{0.48\textwidth}
		\centering
		\includegraphics[width=\linewidth]{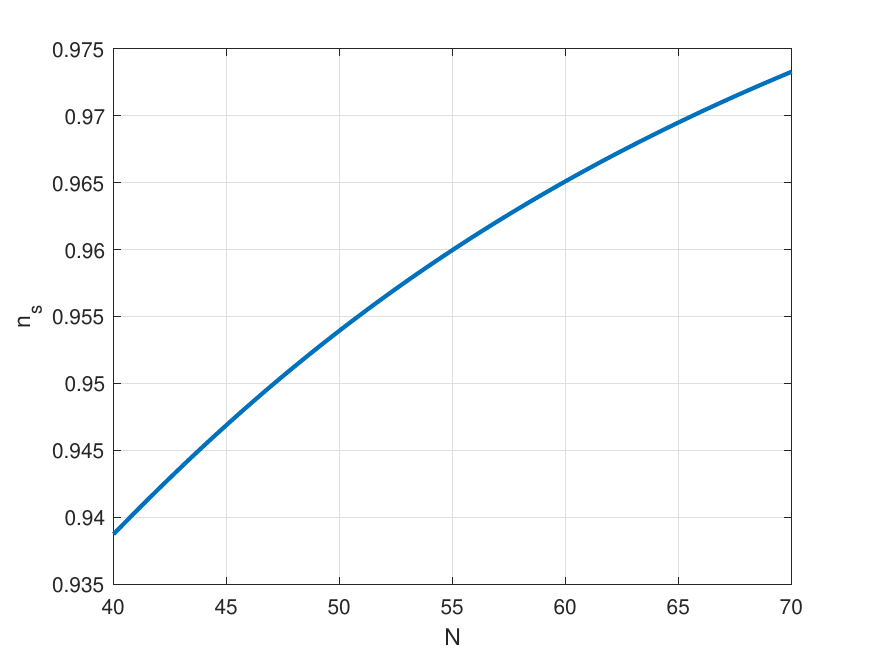}
	\end{subfigure}
	
	\caption{
			Numerical evolution of the inflationary observables. 
			Left panel: tensor-to-scalar ratio $r$ as a function of the number of e-folds $N$. 
			Right panel: scalar spectral index $n_s$ as a function of $N$. 
			The parameters are fixed to $\xi = 0.0083$, $\gamma = 1$, $\kappa = 1$, and $C_4 = 1$.
	}
	\label{figure_1_cos}
\end{figure}

\begin{figure}[htbp]
	\centering
	\includegraphics[width=0.7\textwidth]{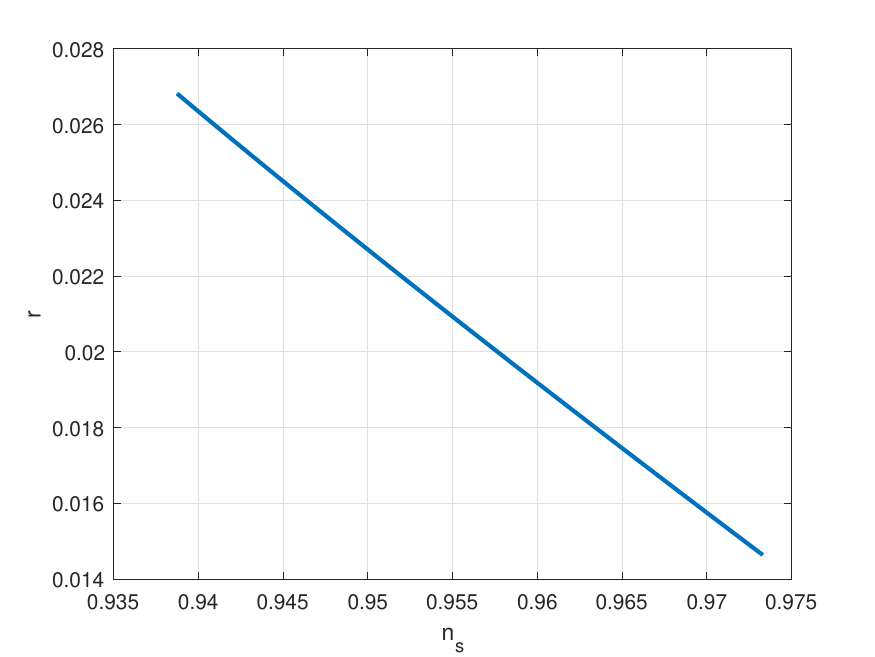}
	\caption{
			Tensor-to-scalar ratio $r$ as a function of the scalar spectral index $n_s$.
			The parameters are fixed to $\xi = 0.0083$, $\gamma = 1$, $\kappa = 1$, and $C_4 = 1$.
	}
	\label{figure_2_cos}
\end{figure}

\clearpage

In the framework of the modified $f(Q,\phi)$ gravity, we focus here on the specific case of the \textit{Cosh-type inflationary model} and analyze the behavior of the inflationary observables $r$ and $n_s$ as functions of the number of e-folds $N$. The results are displayed in Fig. \ref{figure_1_cos}, \ref{figure_2_cos}.

The left panel shows that the tensor-to-scalar ratio $r(N)$ is a monotonically decreasing function of $N$. More precisely, $r$ decreases from $r \sim 2.7 \times 10^{-2}$ at $N = 40$ to $r \sim 1.4 \times 10^{-2}$ at $N = 70$. This behavior is characteristic of a smooth slow-roll evolution, where tensor perturbations are progressively suppressed as inflation proceeds. In the context of the Cosh-type potential embedded in $f(Q,\phi)$ gravity, this suppression is enhanced by the nonminimal coupling between the scalar field and the nonmetricity scalar $Q$.

Conversely, the scalar spectral index $n_s(N)$ exhibits a monotonically increasing behavior. It evolves from $n_s \sim 0.94$ at $N = 40$ to $n_s \sim 0.973$ at $N = 70$, approaching the nearly scale-invariant regime. This feature reflects the ability of the Cosh-type inflationary potential, within the $f(Q,\phi)$ framework, to generate stable scalar perturbations consistent with observational requirements.

The parametric representation $r(n_s)$ further reveals a clear inverse correlation between these two observables: as $n_s$ increases, $r$ decreases. This trend is consistent with standard single-field inflationary predictions, but here it is modulated by the geometric contribution of the nonmetricity through the $f(Q,\phi)$ coupling.

Focusing on the observationally relevant value $N = 60$, we obtain
\begin{equation}
n_s \approx 0.965 - 0.967, \qquad r \approx 0.017 - 0.018,
\end{equation}
which are fully compatible with the latest Planck and BICEP/Keck constraints. In particular, the predicted scalar spectral index lies within the $1\sigma$ confidence interval, while the tensor-to-scalar ratio remains safely below the current upper bound $r < 0.06$.

These results demonstrate that the Cosh-type inflationary scenario within the $f(Q,\phi)$ gravity framework provides a viable and robust description of early-universe inflation. Moreover, the nonminimal coupling plays a crucial role in controlling the dynamics, allowing the model to naturally satisfy observational constraints while maintaining a well-behaved inflationary evolution.

\section{Conclusion}

In this work, we have investigated inflationary dynamics within the framework of modified $f(Q,\phi)$ gravity, where a scalar field is nonminimally coupled to the nonmetricity scalar. This framework provides a natural geometric extension of standard scalar-field inflation and allows for a richer phenomenology driven by the coupling parameter $\xi$.

We have analyzed in detail two representative inflationary scenarios, namely the De Sitter and Cosh-type models, and examined the behavior of the main inflationary observables, the scalar spectral index $n_s$ and the tensor-to-scalar ratio $r$. Our results clearly show that the nonminimal coupling plays a crucial role in shaping the inflationary predictions.

In the De Sitter case, we demonstrated that the model is observationally viable only within a restricted range of the coupling parameter. In particular, we found that $n_s$ increases monotonically with $\xi$, while $r$ decreases, leading to a narrow window
\begin{equation}
10^{-3} \lesssim \xi \lesssim 10^{-2},
\end{equation}
within which the predictions are consistent with Planck and BICEP/Keck constraints. Outside this interval, the model either predicts excessively large tensor modes or a blue-tilted scalar spectrum ($n_s > 1$), which is observationally disfavored.

Moreover, theoretical consistency imposes an upper bound on the coupling parameter, $\xi < \frac{\kappa}{2p}$, which for $\kappa = 1$ and $p = 60$ leads to $\xi < 0.00833$. This result is in remarkable agreement with the observationally preferred region, reinforcing the internal consistency of the model.

In contrast, the Cosh-type inflationary scenario exhibits a more stable and robust behavior. The evolution of $r$ and $n_s$ with respect to the number of e-folds $N$ shows a smooth slow-roll regime, characterized by a suppression of tensor modes and a gradual approach to a nearly scale-invariant spectrum. For $N = 60$, the model predicts
\begin{equation}
n_s \approx 0.965 - 0.967, \qquad r \approx 0.017 - 0.018,
\end{equation}
which lie well within current observational bounds. This demonstrates that the Cosh-type model provides a natural and less fine-tuned realization of inflation in $f(Q,\phi)$ gravity.

Overall, our analysis highlights that the coupling between the scalar field and nonmetricity can significantly improve the phenomenological viability of inflationary models. The geometric contributions encoded in $f(Q,\phi)$ effectively modify the dynamics of the scalar field and allow for better control of inflationary observables.

As a perspective, it would be interesting to extend this study by considering more general functional forms of $f(Q,\phi)$, as well as confronting the model with additional observational probes such as reheating constraints, non-Gaussianities, and large-scale structure data. Such investigations could further clarify the role of nonmetricity in early-universe cosmology and provide deeper insights into the fundamental nature of gravity.

\end{document}